\documentclass[aps,pra,preprint,amsmath,amssymb,groupedaddress]{revtex4-1}

\usepackage{graphicx}
\usepackage{color}

\begin{document}

\title{Investigating the electronic properties and structural features of MgH and of MgH$^-$ anions}

\author{L. Gonz{\'a}lez-S{\'a}nchez$^1$, S. G{\'o}mez-Carrasco$^1$, A.M. Santadar{\'i}a$^1$, F.A. Gianturco$^{2}$ \footnote{Corresponding author: francesco.gianturco@uibk.ac.at}, R. Wester$^2$}

\affiliation{$^1$ Departamento de Qu{\'i}mica F{\'i}sica, University of Salamanca, Pza. de los Ca{\'i}dos sn, 37008,  Salamanca, Spain}
\affiliation{$^2$ Institut f\"ur Ionenphysik und Angewandte Physik, Universitaet  Innsbruck, Technikerstr. 25, A-6020, 
	Innsbruck, Austria}

\vspace{10pt}

\date{\today}

\begin{abstract}
 
In the present paper we analyze in detail several properties of the  MgH$^-$ anion and the MgH neutral molecule using accurate ab initio quantum computational methods in order to establish with higher reliability specific molecular features like the gas-phase electron affinity (EA) , the Frank-Condon (FC) factors for excitation of the neutral and of its anion to their lower electronic states, and the general feasibility of employing the anion in photodetachment experiments after its confinement in cold ion traps. The calculations suggest that the EA value is in agreement with an existing early experiment and further places on it a smaller error bar than that given before. Accurate zero-point-energy (ZPE) corrections are also included in our calculations and their effects discussed.

\end{abstract}


\maketitle

\section{introduction}

Molecular anions are known to play an important role in a wide variety of fields: from the chemistry of correlated systems 
\cite{doi:10.1080/00268970902724955,0034-4885-72-8-086401}, to the atmospheric sciences \cite{LI2006179,doi:10.1021/jp711490b} and to the molecular regions of the interstestellar medium (ISM) 
\cite{1538-4357-652-2-L141,fortenberry_2013,1402-4896-1995-T59-035}. 
It still remains rather difficult, however, to investigate 
molecular anions in a controlled manner at ultracold temperatures, while it is now possible to achieve temperatures of a few kelvin by using  beam expansion methods or by examining the trapped particles followed
by buffer gas cooling  \cite{PhysRevLett.101.063201,BIENNIER2014123}. 

It is therefore important to provide such studies of nearly isolated, and cold, molecular anions with a reliable description of the various physical features
that can be revealed in the cold traps observations. Such comparisons nearly always originate from accurate computational studies and therefore it is the purpose of this work to show such test for a specific
polar molecular anion: the MgH$^-$(X$^1\Sigma^+$) case and its neutral counterpart MgH (X$^2\Sigma^+$).

It is interesting to  note here that  the MgH X($^2\Sigma^+$) molecule has been detected a while ago in stellar atmospheres using its optical spectrum \cite{1945ApJ...102..154B,1980ApJ...235..925T}, 
while no observation for it has as yet been made 
in Molecular Clouds or Dark Cores of the interstellar regions. Furthermore, its negative ion has not been observed in any of the Circumstellar Envelopes (CSEs) where most of the non-metallic anions have
been observed over the years (e.g. see: Millar {\it et al} \cite{doi:10.1021/acs.chemrev.6b00480}). It is therefore important to help to support future possible observational searches by being able to assign specific electronic transitions using accurate calculations of the states involved and to provide specific
indications to photodetachment experiments of small molecular ions as those \cite{0004-637X-776-1-25,doi:10.1021/jz402264n} which have been already analysed in the recent literature on small molecular polar anions.

It is also worth mentioning here that the experimental studies of molecular anions, like OH$^-$,  have been able to selectively photodetach the extra electron from specific rotational states of the relevant target by first knowing both the EA and the rotational 
constant values with substantial accuracy \cite{Endres2017134,Hauser-etal:15}.

The MgH$^-$ anion, however, has received thus far comparatively less attention after its first observation, nearly fifty years ago, when it was generated in a Penning discharge negative ion source \cite{BETHGE1966542}. Later experiments on its photodetachment process had provided the first assessment of its EA value \cite{1977ZNatA..32..594R}. 
The most recent experiments were carried out on both MgH$^-$ and MgD$^-$ using yet another type of negative ionic source, a pulsed and cluster ionization source (PACIS) \cite{Li356},
and crossing a mass-selected beam of negative ions 
with a fixed-frequency photon beam \cite{BUYTENDYK2014140}. These experiments were therefore able to determine the EA values to be, respectively, 0.90$\pm$0.05 eV and 0.89$\pm$0.05 eV., i.e. with rather large error bars. They also carried out ab initio calculations at the
CCSD(T)/aug-cc-pVQZ level of theory and found calculated EA values of 0.86 and 0.85 EV, respectively. The earlier measurements of \cite{1977ZNatA..32..594R} had given an EA value for MgH$^-$ of 1.05 eV, which is higher than the 
later experiments of \cite{BUYTENDYK2014140}. There is therefore a level of uncertainty as to which value should be selected from experiments, 
although no earlier data exist for MgD$^-$. Later calculations of the EA value for MgH$^-$ were carried out by \cite{CHEM:CHEM200801012} and reported also a different value for it: 0.83 eV.

In the next Section we briefly report our computational approach, while the calculated results for MgH$^-$ will be given in Section 3. Our present findings will be finally discussed in Section 4.

\section{Ab initio calculations}

Ground state and several excited state potential energy curves have been obtained for the MgH$^-$ anion and the MgH neutral molecules. Electronic calculations have been performed using the
MOLPRO \cite{MOLPRO_brief} suite of {\it ab initio} quantum chemistry code. Correlation-consistent polarized  valence quintuplet zeta basis sets, denoted as aug-cc-pV5Z and aug-cc-pwCV5Z, have been used for hydrogen and magnesium, respectively \cite{doi:10.1063/1.4865749}. All calculations have been done in the $C_{2v}$ point group of symmetry. 

In the case of the anion, we have focussed on the electronic states correlating with  Mg($^1S$)+H$^-$($^1S$) and Mg$^-$($^2P$)+H($^2S$). A state averaged complete active space calculation (SA-CASSCF) has been initially selected. We have further tried different active spaces and selected the one that correctly describes all electronic states of interest, correlating with the above mentioned asymptotic limits. The active space has finally consisted of all the configurations arising from distributing 4 electrons in 19 orbitals. 
The number of states included in the SA-CASSCF has been selected from the asymptotic limits, adding other excited states which were becoming important at specific geometries. In total, two $^1\Sigma^{+}$, two $^3\Sigma^{+}$, one $^1\Delta$, one $^1\Pi$ and one $^3\Pi$ have been included. After that, an internally-contracted multi-reference configuration-interaction method (icMRCI) has been performed, including the Davidson correction, which estimates the contribution of higher excitation terms. For the neutral MgH molecule, we have calculated the electronic states correlating with Mg($^{1,3}S, ^{1,3}P$)+H($^2S$) and the same basis sets and method have been used. In this case, the number of states included in the state averaged has been six $^2\Sigma^{+}$,  three $^2\Pi$ and one $^2\Delta$ \cite{cpl:2010}.

All the bound ro-vibrational energy levels of the different states and the Franck-Condon factors between
different vibrational states, were calculated using the LEVEL16 program 
\cite{LeRoy2017167}. By including the zero point vibrational energy, the electron affinity was finally evaluated as we will further discuss in the next Section.

\section{Results from computations}

\begin{table}
	\caption{\label{tab:table1}Computed properties of the neutral MgH and anionic MgH$^-$ ground electronic states,
	as a function of the increasing quality level of the chosen basis set expansion.}
	\begin{ruledtabular}
		\begin{tabular}{ccccc}
			& 3Z\footnotemark[1] & 4Z\footnotemark[2] & 5Z\footnotemark[3] & CBS \\
			 \hline
			{\bf	MgH$^-$ (X$^1\Sigma$)} &  &  &  &  \\
			Eq. dist. / \AA & 1.8650 & 1.8651 & 1.8653 & 1.8654 \\
			D$_e$ / cm$^{-1}$ & 12365.5 & 12392.7 & 12397.5 & 12400.2 \\
			ZPE / cm$^{-1}$ & 556.4 & 556.3 & 556.0 & 555.8 \\
			B$_0$ / cm$^{-1}$ & 4.8993 & 4.8988 & 4.8978 & 4.8972 \\
			$\omega_e$ / cm$^{-1}$ &  1130.31 & 1130.15 & 1129.43 & 1129.05 \\       
			\hline
			{\bf	MgH (X$^2\Sigma$)} &  &  &  &  \\
			Eq. dist. / \AA & 1.7436 & 1.7442 & 1.7436 & 1.7432 \\
			D$_e$ / cm$^{-1}$ & 11531.6 & 11527.2 & 11524.9 & 11523.6 \\
			ZPE / cm$^{-1}$ & 747.1 & 729.5 & 748.0 & 759.0 \\
			B$_0$ / cm$^{-1}$ & 5.6539 & 5.6510 & 5.6538 & 5.6555 \\
			EA / cm$^{-1}$ & 7072.1 & 7078.7 & 7104.7 & 7119.9 \\
		 	$\omega_e$ / cm$^{-1}$ &  1508.49 & 1466.90 & 1510.68 & 1535.68 \\       
		\end{tabular}
	\end{ruledtabular}
	\footnotetext[1]{H = aug-cc-pV5Z, Mg = aug-cc-pwcVTZ}
	\footnotetext[2]{H = aug-cc-pV5Z, Mg = aug-cc-pwcVQZ}
	\footnotetext[3]{H = aug-cc-pV5Z, Mg = aug-cc-pwcV5Z}
\end{table}

\begin{table}
	\caption{\label{tab:table2}Computed properties of the lowest four excited electronic states of the anionic MgH$^-$ molecule, with $5Z$ level of expansion.}
	\begin{ruledtabular}
		\begin{tabular}{cccccc}
			MgH$^-$ &  Eq. dist. / \AA & D$_e$ / cm$^{-1}$ & ZPE / cm$^{-1}$ & B$_0$ / cm$^{-1}$ & $\omega_e$ / cm$^{-1}$ \\
			\hline
			{\bf	(a$^3\Pi$)} &   1.7622 & 12904.7 & 689.1 & 5.5172 & 1388.89 \\
			{\bf	(A$^1\Pi$)} &  1.7501 & 10583.8 & 719.1 & 5.6108 & 1445.66 \\
			{\bf	(b$^3\Sigma$)} &  1.7459 & 9920.7 & 707.2 & 5.6304 & 1424.34 \\
			{\bf	(B$^1\Sigma$)} & 1.7368 & 9461.6 & 736.5 & 5.6988 & 1484.32 \\
		\end{tabular}
	\end{ruledtabular}
\end{table}

In order to asses the extent of reliability that can be placed on the present study, we report on the following Table \ref{tab:table1} and Table \ref{tab:table2}   some of the qualifying properties of the neutral molecule, MgH, and of its anion. 
The data are presented as a function of increasing  quality of the basis set employed, up to the complete basis set (CBS) extrapolation which we have employed. For completness we also present in that table the calculations for the rotational constant and of the  vibrational frequency in its harmonic approximation.

We clearly see that  convergence of the properties examined is already achieved at the 5Z level of expansion, since the extrapolation to the CBS values changes them only  marginally.
In the case of the EA values (ZPE corrected), we see that it changes by 15 cm$^{-1}$ from the 5Z value, i.e. about 0.002\%. For the calculations of the other properties reported below, we
therefore decided to stay at the 5Z level of basis set expansion for our additional calculations discussed below.

\begin{figure}[h]
\includegraphics{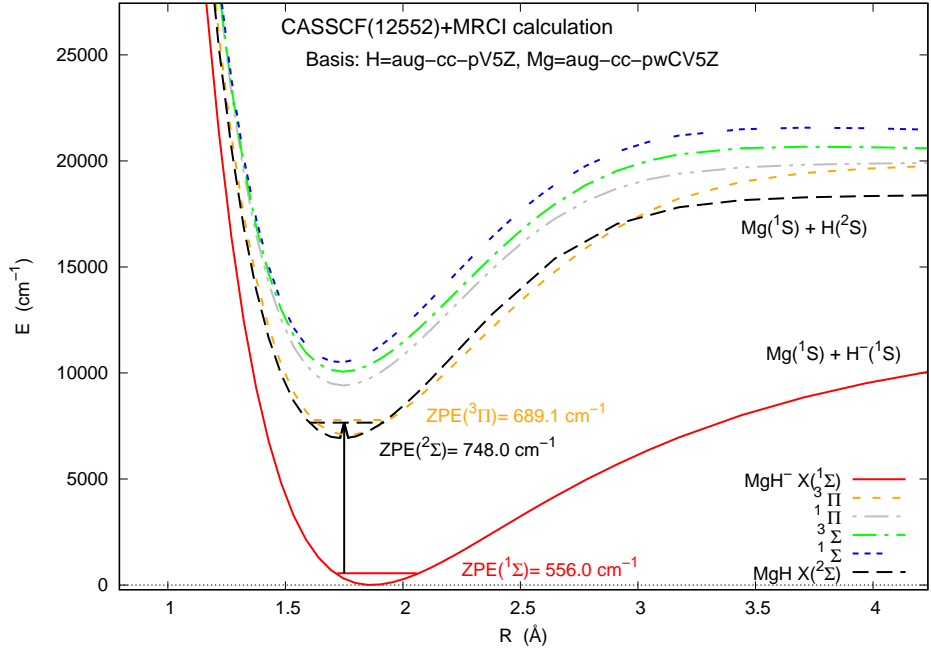}
	\caption{\label{fig:fig1}Pictorial view of the lower electronic potential energy curves for the MgH and MgH$^-$ systems. The ZPE values are given for the lowest three electronic potentials. See main text for further details.}
\end{figure}

A pictorial view of the relevant Born-Oppenheimer (BO) potential energy curves is reported by the Figure \ref{fig:fig1}, where we also indicate the zero-point-energy (ZPE) values obtained by numerical 
integration of the relevant potentials for each electronic state considered \cite{LeRoy2017167}.

The following comments can be made from a perusal of the results shown in that figure:
\begin{enumerate}
	\item Once we calculate the EA value between the two lowest electronic states for the neutral and the anion (see values in Table \ref{tab:table1}) and correct it with the ZPE differences between the two
		potential energy curves (PECs), we find a value of 7104.7 cm$^{-1}$, corresponding to about 0.881 eV. Our calculated value is expected to carry a numerical error of at most 15 cm$^{-1}$, as mentioned earlier. The latest
		experimental assessment \cite{BUYTENDYK2014140} is 7258.0 cm$^{-1}$, with the error of $\pm$403.3 cm$^{-1}$. We can  therefore surmise that the present extrapolated value can reduce the error bar on the latest experiments by almost 130-140 cm$^{-1}$ and place the best available EA estimate to be around 7120 cm$^{-1}$, corresponding to about 0.88 eV, which is well with the
		experimental error bar of ref. \cite{BUYTENDYK2014140} (0.90$\pm$0.05 eV).
	\item The calculations further show that the first electronic excited state of the MgH$^-$, the ($^3\Pi$) state, dissociates above the ground electronic state of the neutral, thereby indicating its metastability and 
		therefore the possibility of its undergoing autodetachment from the vibrationally excited states of that anion. This feature has been noted before for various diatomic anions \cite{PhysRevLett.114.213001}, thus suggesting that it might be
		difficult for such states to undergo Doppler cooling  of the anions in cold traps \cite{PhysRevA.89.043410} , thereby favouring the cooling path via buffer gas interaction with, for example, He atoms.
	\item In the geometry regions around the v=0 levels for both MgH ($^2\Sigma^+$) and the anion's excited state ($^3\Pi$) we further see that the anion's level lies above, albeit by a very small energy
		difference, the same level of the neutral. This feature indicates a metastability region for the anion which can then  undergo  electron autodetachment and stabilize the v=0 levels of the neutral 
		partner.
\end{enumerate}

\begin{figure}[t]
	\includegraphics[height=0.6\textheight]{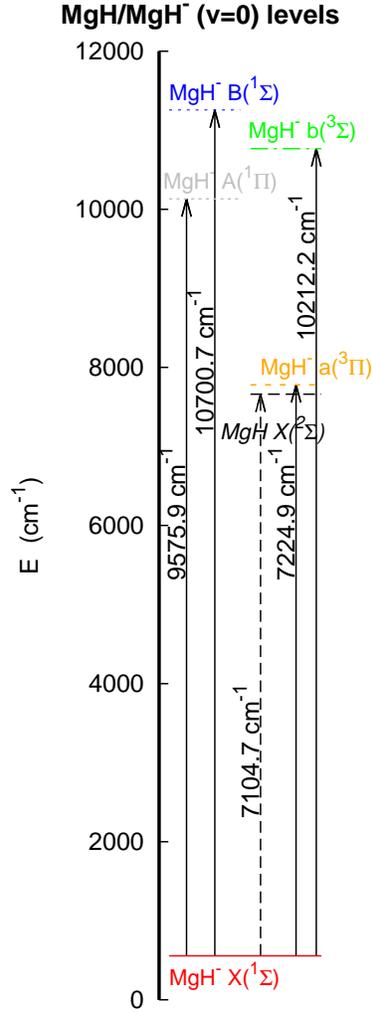}
	\caption{\label{fig:fig2}Calculated transition energies (at the 5Z level of basis set choice) from the ground electronic state of the MgH$^-$ anion and of its neutral counterpart. On the left side of the figure we show the transition energies
	to the 2nd and 4th excited electronic state of the anion, while on the right we show the excitations to two more excited states, the 1st and the 3rd states. We show also on that side of the figure
	the electron-detachment process to final  MgH($^2\Sigma$) neutral product. See main text for further details.}
\end{figure}

  To  further illustrate the behaviour of the excitation transitions between electronic states of the two molecular systems of the present study, we report in the next Fig. \ref{fig:fig2} all the values between v=0 levels (ZPE corrections included) of the electronic states given by Figure \ref{fig:fig1}.

One interesting feature of the figures is the marked proximity in energy between the final state for the photodetachment process (the MgH X($^2\Sigma$) state) and the one for the first electronic excitation 
of the anion: the a($^3\Pi$) state of the MgH$^-$. Furthermore, as already mentioned , the negative ion state lies above the one of the neutral (ZPE corrections are included). This indicates a possible
metastability for the excited electronic state of the anion, which could then decay to the neutral ground electronic state and eject the excess electron.

Another important property that can tell us of the transition efficiency between the considered levels is obtained by generating the  Franck-Condon (FC) overlap integrals between the lower vibrational levels supported 
by the computed electronic PECs. The values obtained from our calculations are reported by the next Table \ref{tab:table4}. We show there two different processes: the upper panel reports the photodetachment of the excess electron of the anionic molecule from three of its lower vibrational states,  with the formation of the ground electronic state of the neutral MgH final product into three different  vibrational levels.

The lower part of the Table \ref{tab:table4} shows instead different electronic excitation processes from the ground electronic state of the anion, the $^1\Sigma^+$ state, into four different excited electronic states of the same
anion. One clearly sees in the table that the transitions associated with the partners being in their ground vibrational states are markedly favoured and are associated with  larger FC factors, the largest being
, in relative terms, the one associated with  the $\left| ^1\Sigma^+ \right> \leftarrow \left| ^3\Pi \right>$ transition.

Naturally, other factors come into controlling the size of their transition moments, e.g. change of electronic angular momenta and also change of the electronic spin multiplicity. However, the calculations indicate that 
to have the molecular ions in their ground vibrational levels will provide the most favourable FC factors for the corresponding transitions.

Since both the neutral molecule and its corresponding negative ions are  polar molecules, to have specific, and reliable, knowledge of their permanent dipole moments is clearly an important factor for
the present discussion. The data for the neutral species have been in the literature for a while \cite{doi:10.1063/1.1673619,doi:10.1080/00268979100101041}, 
with recent analysis being carried out over a broad range of molecular geometries \cite{GUITOU2010145,doi:10.1111/j.1365-2966.2012.21367.x}, 
including a molecular line opacity study in cold stellar atmospheres \cite{0004-637X-584-1-459}. The electric dipole moment of MgD has also been experimentally studied \cite{doi:10.1063/1.4878414} 
and a value of 1.318 D found for its (X $^2\Sigma$)
state, with a marked increase up to 2.567 D for the (A $^2\Pi$) excited electronic state. Nothing, to our knowledge, exists however for the dipole moment of its anionic counterpart.

The calculations which we report in Table \ref{tab:table4} show the variations of the computed dipole moment values as a functions of the size of the basis sets employed: they are in the same sequence which we have already discussed earlier on in Tables \ref{tab:table1} and \ref{tab:table2}.

We clearly see that the negative ion consistently presents a larger value for its permanent dipole moment, as should be expected by the presence of the additional negative charge along the bond. By
the time we employ the largest basis set discussed earlier, we also see that the value known for MgD, which is expected to be smaller than that of MgH, indicates -1.318 D at its equilibrium geometry \cite{doi:10.1063/1.4878414}.
This is suggesting that our estimate of the dipole moment for the MgH partner is fairly reliable for the equilibrium geometry which we found in our calculations. We also report in the next Figure \ref{fig:fig3} our calculated dipolar function for the ground  electronic state of MgH and compare it with earlier calculations for the same quantity.

\begin{table*}
	\caption{\label{tab:table3}Computed FC factors for the photodetachment process from the ground electronic state of MgH$^-$ (upper panel)
	and for the electronic excitation of MgH$^-$ to its four electronic states (lower panel).}
	\begin{ruledtabular}
		\begin{tabular}{lccccccc}
			\multicolumn{4}{l}{{\bf Photodetachment from MgH$^-$(X$^1\Sigma^+$,v'')}}&\multicolumn{4}{l}{}\\
			 & \multicolumn{3}{c}{v''=0} & &\multicolumn{3}{c}{v''=1} \\
			 & v'=0 &v'=1 & v'=2 & & v'=0 &v'=1 & v'=2 \\
			\multicolumn{4}{l}{Final state: MgH($^2\Sigma$,v')}& \multicolumn{4}{l}{}\\
			FC & 0.725 & 0.235 & 0.036 & &0.220 & 0.298 & 0.363 \\
			Energy gap / cm$^{-1}$ & 7104.7 &8554.4& 9936.1& &6050.2 & 7498.9 & 8879.5 \\
			\hline
			\hline
			\multicolumn{4}{l}{{\bf Excitations from MgH$^-$(X$^1\Sigma^+$,v'')}}&\multicolumn{4}{l}{}\\
			 & \multicolumn{3}{c}{v''=0} & &\multicolumn{3}{c}{v''=1} \\
			 & v'=0 &v'=1 & v'=2 & & v'=0 &v'=1 & v'=2 \\
			\multicolumn{4}{l}{Final state: MgH$^-$($^3\Pi$,v')}& \multicolumn{4}{l}{}\\
			FC & 0.808 & 0.177 & 0.014 & &0.162 & 0.469 & 0.315 \\
			Energy gap / cm$^{-1}$ & 7224.9 &8560.4& 99843.3& &6164.5& 7500.0 & 8782.9\\
			\hline
			\multicolumn{4}{l}{Final state: MgH$^-$($^1\Pi$,v')}& \multicolumn{4}{l}{}\\
			FC & 0.754 & 0.217 & 0.027 & &0.200 & 0.352 & 0.355 \\
			Energy gap / cm$^{-1}$ & 9575.9 &10974.0&12313.6& &8515.5& 9913.6 & 11253.2\\
			\hline
			\multicolumn{4}{l}{Final state: MgH$^-$($^1\Sigma$,v')}& \multicolumn{4}{l}{}\\
			FC & 0.703 & 0.253 & 0.040 & &0.230 & 0.263 & 0.376 \\
			Energy gap / cm$^{-1}$ & 10700.7&12132.7&13505.4& &9640.3&11072.3 & 12446.4\\
			\hline
			\multicolumn{4}{l}{Final state: MgH$^-$($^3\Sigma$,v')}& \multicolumn{4}{l}{}\\
			FC & 0.746 & 0.226 & 0.027 & &0.203 & 0.341 & 0.365 \\
			Energy gap / cm$^{-1}$ & 10212.2&11587.1&12908.1& &9151.8&10526.7 & 11847.7
		\end{tabular}
	\end{ruledtabular}
\end{table*}

\begin{table}
	\caption{\label{tab:table4}Computed and measured permanent dipole moments for the lowest electronic states for the MgH neutral molecule
	and its anion. The upper  panel shows the variations as a function of basis set qualities. The bottom panel reports the experimental value
	at the equilibrium geometry of the neutral. See main text for further details.}
		\begin{tabular}{lccccc}
			\hline
			\hline
			& & &\multicolumn{3}{c}{Dip. mom. / Debye }\\
			& r / a$_0$& & 3Z& 4Z & 5Z  \\
			MgH (X$^2\Sigma$) & 3.300& & -1.366 & -1.365 & -1.364\\
			MgH$^-$ (X$^1\Sigma^+$) & 3.500& & 1.854 & 1.867 & 1.864 \\
			\hline
			\multicolumn{6}{c}{MgH (X$^2\Sigma$) } \\
			Ref. & r$_{eq}$ / a$_0$& &\multicolumn{3}{c}{Dip. mom. / Debye}\\
			\cite{doi:10.1063/1.1673619} & 3.271 & &\multicolumn{3}{c}{-1.511} \\ 
			\cite{doi:10.1111/j.1365-2966.2012.21367.x} & 3.269 & &\multicolumn{3}{c}{-1.371} \\ 
			\hline
			\hline
		\end{tabular}
\end{table}

\begin{figure}[h]
	\includegraphics{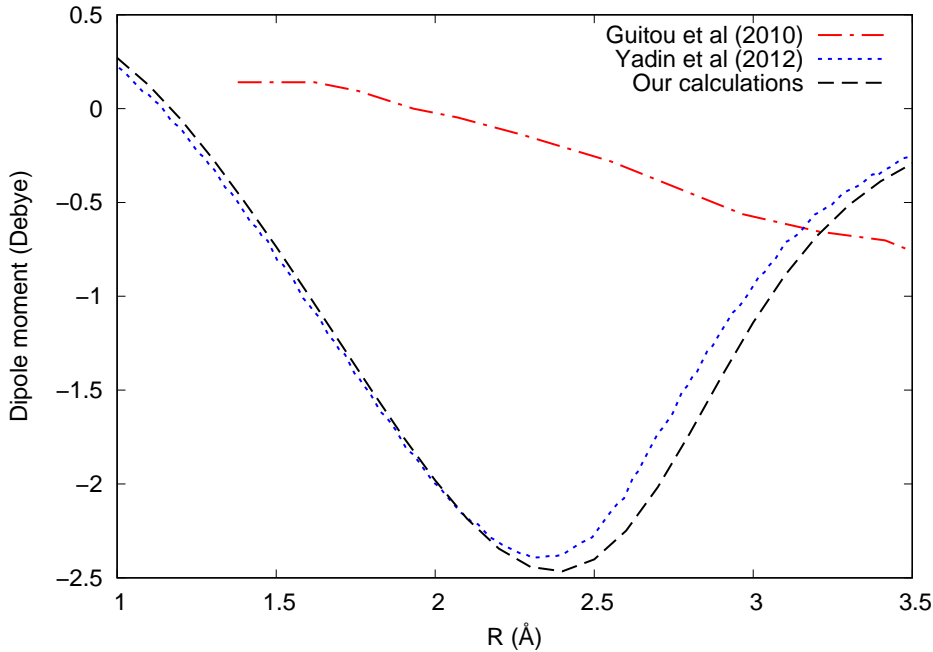}
	\caption{\label{fig:fig3}Computed dipolar function for the {\color{red} MgH (X$^2\Sigma$)} electronic state for a range of geometries from 1.0 to 3.5 \AA. Two further calculated functions are shown for comparison. See main text for further details.}
\end{figure}

We see from the data in the figure that the $\mu$ value which we have calculated decreases with distance around the region of the equilibrium geometries while increasing as the bond is either stretched and/or compressed. by also looking at our results of Table \ref{tab:table3} we see that the earlier estimates from
ref \cite{doi:10.1063/1.1673619} follow very closely our computed behaviour in that Figure while the more recent data from \cite{GUITOU2010145,doi:10.1111/j.1365-2966.2012.21367.x} 
are exhibiting a very different dependence on the changing of the internuclear distances.

 We can therefore argue from such data that both MgH and MgH$^-$,
in their respective ground electronic states, have substantial dipole moment values but are still too small to reach the critical  value which would suggest the presence of dipole-bound (DB) configurations associated with very small binding energies for the diffuse excess electron of that type of anion \cite{doi:10.1063/1.1358863}. This is a important result in relation to our discussion on the properties of the metastable anion that we shall further discuss in the next Section.

\section{Present conclusions}

We carried out very accurate ab initio calculations involving some of the electronic properties of the MgH molecule in the gas-phase and of its corresponding anion, MgH$^-$. The task was to assess the feasibility of carrying out photodetachment experiments on this system, having experimentally  prepared it in a selected rovibrational state by a cooling procedure in an ion trap, in analogy with similar studies carried out recently in our group on the
OH$^-$ system \cite{Endres2017134,Hauser-etal:15}.

The present calculations have employed a range of basis set expansion in order to assess the overall reliability of the final data of this study, which have been also compared with existing, earlier calculations and 
measurements.

Thus, our calculations have been able to assign a smaller error bar on the value of the EA of the MgH(X$^2\Sigma^+$) and to correct it with the inclusion of ZPE effects and adiabatic corrections. The earlier measured data carried a much
larger error bar and therefore we feel that our present calculations have provided an overall better value for such quantity.

We have also analysed the behaviour of the permanent dipole moments of both MgH(X$^2\Sigma^+$) and MgH$^-$(X$^1\Sigma^+$), finding good agreement for the latter molecule with the most recent determinations, both from 
theory \cite{doi:10.1111/j.1365-2966.2012.21367.x} and experiments \cite{doi:10.1080/00268979100101041}. 
Both molecules turn out to have subcritical dipole moment values and therefore cannot support DB anionic states, which would be associated with very small EA values.

We have also analysed the relative locations of the lowest, excited electronic states of both molecules and found that the next electronic state of MgH$^-$ is an (a$^3\Pi$) excited state which lies by 
about 120 cm$^{-1}$ above  the ground electronic state of the neutral, MgH(X$^1\Sigma^+$). This is an interesting result which suggests  the possible presence of a metastable anion within the photodetachment continuum.
During the actual laser-induced photodetachment experiments \cite{0004-637X-776-1-25} on the present molecular anion one can therefore expect the presence of a Feshbach resonance near the continuum threshold which could affect the
lifetimes of the interacting partners during the electron emission mechanism. In other words, one could actually be able to see a marked signal above threshold corresponding to the formation of the metastable anion as a 
Feshbach resonance. One should also keep in mind, however, that the spin-flip involved in this excitation might strongly reduce the probability of it being visible experimentally. Further electron scattering studies would therefore be needed to provide additional computational evidence. 

The present calculations indicate that MgH$^-$, although studied very little thus far in the literature, could be a good candidate for a polar molecular negative ion which would be amenable to cold trap experiments,
as already suggested recently \cite{PhysRevLett.114.213001}. Thus, its photodetachment analysis from a preselected initial rovibrational state can provide specific indications on the relevant transition moment and on the
possible role played by its metastable excited electronic state which has been found by the present calculations.

Since the selective preparation in a given initial state requires the usage of He gas as a buffer gas in the ion trap \cite{Endres2017134,Hauser-etal:15}, we are currently carrying out the calculation of the full potential energy surface (PES)
for the MgH$^-$(X$^1\Sigma^+$) interacting with He($^1S$) atoms in order to model the quantum dynamics of selective rotational cooling of the trapped ion. The results of this additional study will be reported in a separate paper in the near future.

\begin{acknowledgments}

F.A.G. and R.W. thank the FWF (Austrian Science Fund) for supporting the present research through Project No. P27047-N20. L.G.S. acknowledges funding by the Spanish Ministry of Science and Innovation Grants No.
CTQ2012-37404-C02, CTQ2015-65033-P, and Consolider Ingenio 2010 CSD2009-00038.

\end{acknowledgments}

\end{document}